\documentclass[twocolumn,showpacs,preprintnumbers,amsmath,amssymb]{revtex4}
\usepackage{graphicx}
\usepackage{dcolumn}

\begin{document}
\preprint{}

\title{First-principles study of the structural energetics of PdTi and PtTi} 

\author{Xiangyang Huang, Karin M. Rabe and Graeme J. Ackland}
\affiliation{
Department of Physics and Astronomy, Rutgers University, 
Piscataway, New Jersey 08854-8019}

\date{\today}
\begin{abstract}

The structural energetics of PdTi and PtTi have been studied using
first-principles density-functional theory with pseudopotentials and a
plane-wave basis.  We predict that in both materials, the experimentally
reported orthorhombic $B19$ phase will undergo a low-temperature phase
transition to a monoclinic $B19'$ ground state.  Within a soft-mode framework, 
we relate the $B19$ structure to the cubic $B2$ structure, observed
at high temperature, and the $B19'$ structure to $B19$ via phonon modes 
strongly coupled to strain.  In contrast to NiTi, the $B19$ structure is 
extremely close to hcp. We draw on the analogy to the bcc-hcp transition 
to suggest likely transition mechanisms in the present case.

\end{abstract}


\maketitle

\section{Introduction}

Shape memory alloys (SMA) have attracted a great deal of attention due to
their important technological applications, including mechanical actuator
devices and medical stents.  The shape memory effect also gives rise to 
superelasticity, which finds applications in stents and spectacle frames. 

The shape memory effect is related to a reversible martensitic 
(diffusionless) phase transformation. It has been shown
that the martensitic transformation can be induced by applied fields,
temperature or both, and the mechanical properties of materials,
therefore, can be controlled accordingly.  
In many systems, including those discussed in the present work,
alloying can dramatically
change the properties and transition temperatures of the materials,
reflecting the importance of electronic features, specifically
Fermi surface effects, in the structural energetics of SMA.

There are several complementary approaches to 
modelling of the shape memory effect. Continuum modelling
allows investigation of the microstructural behavior, specifically
martensitic twins, at
the relevant long length scales. Material-specific behavior
is incorporated through an empirical functional for the free energy
in terms of strain and a twin boundary energy to set the 
length scale. In atomistic models, the individual atoms are considered
explicitly and their interaction given by an  
interatomic potential, which may be determined empirically, from 
first-principles density-functional-theory (DFT) calculations, or a 
combination of the two. Crystal symmetry and defect energies emerge 
from this approach, which gives microstructures with both natural length scales
(from defect energies) and time scales (since the atoms have definite 
Newtonian forces and masses). However, in atomistic models, the electronic degrees
of freedom do not appear explicitly. First principles DFT methods are 
so computationally intensive that direct studies of microstructural
behavior are impossible, but they are valuable both for obtaining quantitative
atomic-level information regarding energies, forces and stresses
independent of empirical input, and for understanding the electronic
origin of this behavior. Thus, first-principles investigation of
the energetic instability of the high-temperature structure towards the
low-symmetry martensitic structure is in itself quite illuminating.
The resulting information can then also be used as inputs
to atomistic\cite{Pinsook} and continuum modelling of shape memory
behavior. 

Typically, martensitic transformations are described using the strain
as an order parameter, the classic example being the Bain bcc-fcc
transformation of iron.  However, there is an alternative approach
appropriate for cases where the strain degrees of freedom are coupled to 
atomic degrees of freedom (phonons). Following the soft-mode theory of structural
transitions,\cite{Cochran:1960} we start from a high-symmetry reference
structure (here $B2$) and freeze in unstable phonons of this structure,
with corresponding lattice relaxation, to
produce the ground-state structure.  The symmetry of the phonons 
determines the symmetry of the low temperature structure. 
This approach has been successfully used in the study of 
minerals\cite{GJA,MCW,BBK} and ferroelectric materials
\cite{King-smith:1994,Krakauer:1995,Ghosez:1999,Waghmare:1997} and has
been extended to shape memory alloys in our previous study of
NiTi\cite{Huang:2002}. 

Closely related to NiTi, PdTi and PtTi are shape memory materials with reasonable 
structural simplicity and extraordinary mechanical behavior.
They undergo a martensitic transformation  
at tunable temperatures: PdTi transforms
at 810K, but this can be reduced to 410K with 8\% substitution of 
Cr for Pd\cite{neut}. 
The high-temperature ``austenitic'' phase has a simple cubic $B2$ structure
(space group $Pm3\bar m$),  while the  ambient temperature ``martensitic'' 
phase has been reported as the orthorhombic $B19$ structure\cite{dwight,donkersloot}
(space group: $Pmma$).
Previous first-principles studies in PdTi and PtTi\cite{BEN,Bihlmayer,Ye:1997}
have shown that the observed electronic and elastic properties
of the $B19$ structure are well reproduced by density-functional
theory calculations assuming the experimentally determined structure.

In this paper, we investigate the structural energetics of
PdTi and PtTi from first-principles calculations of phonon frequencies as well as 
total energies. This allows us to examine local as well as global stability and to 
investigate transition mechanisms, drawing on the strong analogy between the 
$B2$-$B19$ and bcc-hcp transformations and showing that coupling of unstable 
modes to the strain is a crucial aspect
of the structural energetics.  In Sec. II, we describe the first-principles
calculations. In Sec. III, we present and discuss the results for the phonon dispersion
of PdTi and PtTi in the $B2$ structure and for the relaxed structures in which unstable 
modes and strains are coupled, yielding a low-symmetry ground state.  In addition, we 
present results of calculations of the electronic structure, identifying and discussing 
features that affect the relative stability of the phases.  Sec. V concludes the paper.

\section{Calculations}

First-principles total energy calculations were carried out within density-functional 
theory with a plane-wave pseudopotential approach. The calculations were
performed with the Vienna {\it ab-initio} Simulations Package\cite{VASP1,VASP2}, 
using the Perdew-Zunger\cite{Perdew:81} 
parametrization of the local-density approximation (LDA).  
Vanderbilt ultrasoft pseudopotentials \cite{Vanderbilt}
were used. Our pseudopotentials include nonlinear 
core corrections and for Ti, we treated the occupied 
$3p$ levels as valence.
The electronic wave functions were represented in a
plane-wave basis set with a kinetic energy cutoff of 278eV.
The Brillouin zone (BZ) integrations were carried out by the 
Hermite-Gaussian smearing technique \cite{Methfessel}
with the smearing parameter of 0.1eV.
The unit cells contain two atoms in the cubic $B2$
structure and four atoms in the orthorhombic $B19$ and monoclinic 
$B19'$ structures.
The calculations were performed with a $16\times16\times16$ 
Monkhorst-Pack (MP) $k-$point mesh for the cubic $B2$ structure and a 
$12\times12\times16$ MP $k-$point
mesh for both orthorhombic $B19$ and monoclinic $B19'$ structures (space group: $P2_{1}/m$),
corresponding to  120 $k$ points in the $1\over48$ irreducible BZ
of the simple cubic cell, 288 $k$ points in the $1\over8$ irreducible BZ of 
the orthorhombic cell and 576 $k$ points in the $1\over4$ 
irreducible BZ of the monoclinic cell.
This choice of parameters converges the total energy to within 
1~meV/atom.
The density of states (DOS) for the $B19$ and $B19'$ structures
were calculated using the tetrahedron method with Bl\"ochl 
corrections\cite{blochl}. 
The BZ's for the orthorhombic $B19$ and monoclinic $B19'$ structures
are different. To compare the band structure of the two structures, 
we label the $B19$ band structure by regarding the $B19$ structure
as a special case of $B19'$.

The phonon dispersion relations were obtained with the PWSCF 
and PHONON codes \cite{SISSAcode}, using
the Perdew-Zunger\cite{Perdew:81} parametrization of 
the LDA, as above.
Ultrasoft pseudopotentials \cite{Vanderbilt}
for Pd, Pt and Ti were generated according to a modified
Rappe-Rabe-Kaxiras-Joannopoulos (RRKJ) scheme \cite{Rappe}
with three Bessel functions \cite{DalCorso}.
The electronic wave functions were represented in a
plane-wave basis set with a kinetic energy cutoff of 408eV.
The augmentation charges were expanded up to 9000eV.
The Brillouin zone (BZ) integrations were carried out by the
Hermite-Gaussian smearing technique \cite{Methfessel} using a 
56 {\it k}-point mesh 
(corresponding to $12\times12\times12$ regular divisions along
the $k_x$, $k_y$ and $k_z$ axes) in the $1\over48$ irreducible
wedge. The value of the smearing parameter was $\sigma$=0.2eV.
These parameters yield phonon frequencies converged within
5 cm$^{-1}$. The dynamical matrix was computed on a
$6\times6\times6$ {\it q}-point mesh commensurate with
the {\it k}-point mesh. The complete phonon dispersion relation was
obtained through the computation of real-space interatomic
force constants within the corresponding box\cite{Giannozzi}.

The choice to use two different first-principles codes was
dictated by the individual strengths of each. VASP has a highly
efficient scheme for calculating total energies, forces, and
stresses, and relaxing to the minimum energy structure, but
does not have the density-functional perturbation theory capabilities
of PWSCF/PHONON. Even with slightly different pseudopotentials and 
{\it k}-point sampling, the results of the two codes are quite compatible.
For example, the difference between the computed lattice parameters for the $B2$
structure of PdTi is less than 0.2\%, and for PtTi the difference
is less than 0.1\%.
Comparisons of normalized eigenvector components computed by VASP using the
frozen phonon method and by PWSCF/PHONON also show good agreement, generally
within 5\%.

\section{Results}

For the cubic $B2$ structure, our calculations yield the
equilibrium lattice parameters of 3.112\AA\ and 3.125\AA\ for PdTi
and PtTi respectively. For comparison, we also performed full-potential
linearized-augmented-plane-wave calculations (FLAPW) within the LDA\cite{FLAPW}. 
The results are in excellent agreement with FLAPW results of
3.113\AA\ (PdTi) and 3.127\AA\ (PtTi) and in 
in good agreement with experimental values of 3.18\AA\ and 3.17\AA.

\begin{figure}
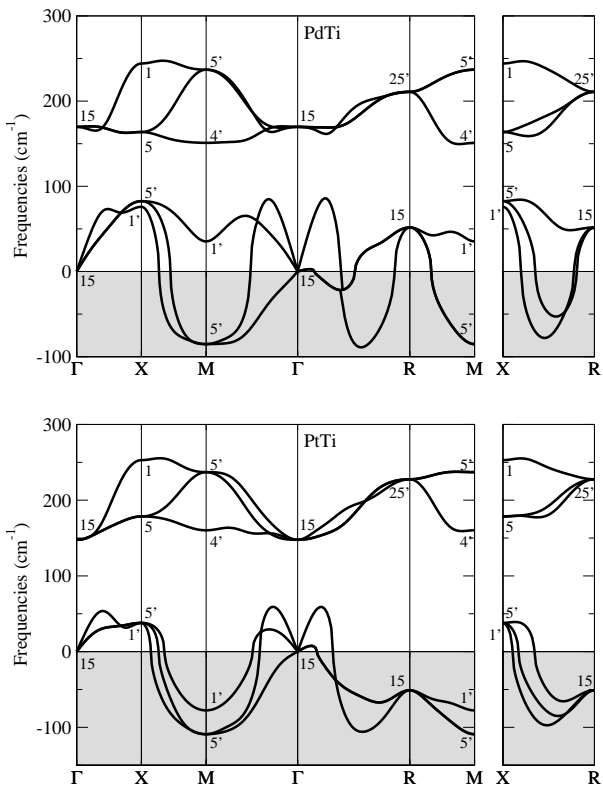

\includegraphics[scale=0.35]{fig-1a.eps}\\[0.5cm]
\includegraphics[scale=0.35]{fig-1b.eps}
\caption{\label{phonons}
Phonon dispersion relations for PdTi (upper) and PtTi (lower) in the $B2$ structure
calculated at the LDA equilibrium parameters 3.112\AA\ and 3.125\AA\, respectively. 
The negative slope of the acoustic $\Gamma-M$ branch corresponds to a pure elastic 
instability ($c'=\frac{1}{2}(c_{11}-c_{12})<0$). Symmetry labels are assigned 
according to the conventions of Ref.~\protect\onlinecite{Bassani} with Pd/Pt 
at the origin. 
The imaginary frequencies of the unstable modes are plotted as negative values.}
\end{figure}

The phonon dispersion relations along high symmetry lines, computed
at the theoretical lattice parameters, 
are shown in Figure~\ref{phonons}. 
The frequencies are obtained by taking the square root of
the eigenvalues of the
dynamical matrix\cite{bornhuang}.
Imaginary frequencies, as plotted as negative,
are obtained from negative eigenvalues of the dynamical
matrix. Thus, the structure is dynamically unstable against distortions
following the corresponding eigenvector.
It should be noted that the phonon frequency $\omega$ is {\it not} the 
reciprocal of the period of oscillation of this mode 
(as measured in molecular dynamics) nor is it the energy difference
between adjacent quantum levels (as measured in neutron scattering 
experiments).  These three quantities are equal only for a stable 
harmonic crystal. In the materials studied here the unstable modes may be 
related to a soft mode\cite{Cochran:1960} 
(as defined by MD or neutron scattering) at high temperature,
and even the stable modes are expected to be significantly
renormalized as a function of temperature by anharmonic effects.

The dynamical matrices are related by mass factors to the force constant matrix: 
the second derivatives of the internal energy with respect to atomic displacements.
The eigenmodes of the force constant matrix describe the potential energy landscape, 
and a negative eigenvalue indicates a static instability against a 
distortion following the corresponding eigenvector. While the
actual normalized displacements of these eigenmodes are in general slightly different
they carry
the same symmetry labels as the eigenmodes of the dynamical matrix.
Either choice is expected to serve as a useful pointer to a lower energy structure if
 the distorted structure obtained by ``freezing in" an unstable mode is 
relaxed using first-principles forces and stresses, as we describe below.

The phonon dispersion relations shown in Figure~\ref{phonons}
show instability of the $B2$ structure similar to and even stronger than that
of NiTi\cite{Huang:2002}. There are large regions of reciprocal 
space where one, two or even three modes are unstable, with dominant 
instabilities at $M$ and along $\Gamma$-$R$. The phonon instability shows 
that the observed high-temperature $B2$ phases of PdTi and PtTi are 
dynamically stabilized by anharmonic phonons, and should be characterized 
by large fluctuating local distortions.
The calculated phonon dispersions are also reminiscent of those of unstable 
bcc materials such as Zr and Ti, which undergo martensitic transformations 
to hcp or $\omega$\cite{Sikka:1982} 
(via the $M$ and $\Gamma$-$R$ bcc-phonon equivalents respectively) phases.  
The analogy based on the view of $B19$ phases of PdTi and PtTi as chemically
ordered hcp will be further strengthened below.

In the soft mode approach, we search for local energy minima by choosing
an unstable mode of the high symmetry structure, freezing in the distortion
with varying amplitude, and relaxing the resulting structure. 
In many cases, the mode with the largest negative eigenvalue will generate
the lowest energy structure. However, this is by no means generally true,
as the energy gain is determined not only by the curvature of the energy
surface but by higher order terms as well as the strength of coupling to
strain and other modes, both unstable and stable, of appropriate symmetry.
Indeed, in PdTi this ``most unstable'' mode
(i.e. largest negative eigenvalue) lies in the $\Gamma$-$R$ branch
rather than at $M$.  This mode is typically unstable in
dynamically-stabilized bcc materials such as titanium and zirconium
where it is associated with a phase transition to the ideal $\omega$ ($C32$)
phase.  
The fact that in the chemically ordered analog (Pd/Pt)Ti, the observed phase 
transition is to $B19$ rather than $C32$
illustrates the importance of anharmonic effects and strain coupling
in the energetics of these materials.

\begin{figure}[bp]
\includegraphics[scale=0.7]{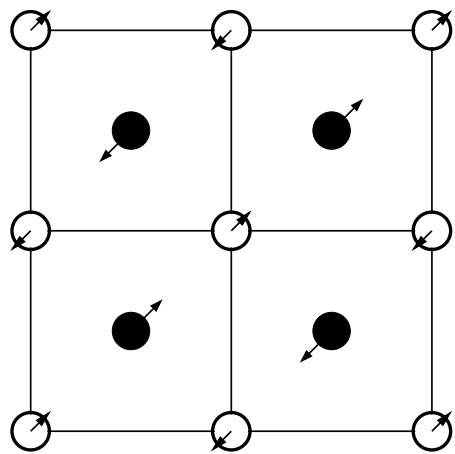}
\caption{\label{pattern}
Eigenmode of the doubly-degenerate $M_5'$ unstable phonon in the $B2$ structure.
This mode generates the $B19$ structure in PdTi and PtTi. The structure is shown 
projected along the $\hat z$ direction, with Pd/Pt represented by filled circles and
Ti by open circles.
}
\end{figure}

\begin{table*}[!htp]
    \caption{\label{table:struct}
     Computed structural parameters and total energies of PdTi and PtTi in the 
     $B19$ and $B19'$ structures, compared with the computed $L1_0$ structure, the computed
     bco structure, the experimental $B19$ structure, with previous 
     calculations\protect\cite{Ye:1997} and with 
     two special $B19$ structures with $b/a$ and $c/a$ corresponding to 
     ideal hcp and $B2$ (bcc) structures. 
     Volume is given in \AA$^3$/formula unit, and energy in eV/atom. 
     Wyckoff positions refer to space group $P2_{1}/m$ ($B19'$).}
    \begin{ruledtabular}
    \begin{tabular}{ccccccccccc}
       & Structure & Volume & $b/a$  & $c/a$ & $\gamma(^{\circ})$ 
        & Wyckoff position & $x$ & $y$ & $z$ & $E-E_{B2}$\\
    \hline
    PdTi & $B19$ 
    & 29.63 & 1.753 & 1.634 & 90    & Pd(2e) & 0   & 0.6866 & 0.25& -0.0917\\
   &&       &       &       &       & Ti(2e) & 0.5 & 0.2008 & 0.25& \\
         & $B19'$ 
    & 29.64 & 1.758 & 1.633 & 93.39 & Pd(2e) &0.0114&0.6827 & 0.25& -0.0924\\
   &&       &       &       &       & Ti(2e) &0.4475&0.1964 & 0.25 \\
         & $L1_0$ 
    & 29.32 & 1.000 & 1.375 & 90    &  &    &    &     & -0.0870\\
         & bco
    & 30.16 & 1.837 & 1.579 & 105.79 & Pd(2e) &0.0882&0.6764 & 0.25& -0.0573\\
   &&       &       &       &       & Ti(2e) &0.3491&0.1983 & 0.25 \\
         & $B19$ \footnotemark[1]
    & 31.33 & 1.74  & 1.62  & 90    &&&&&\\
         & $B19$ \footnotemark[2]
    & 31.74  & 1.75  & 1.64  & 90    &&&&&\\
         & $B19$ \footnotemark[3]
    &  30.3 & 1.72 & 1.62   & 90    & Pd(2e) & 0   & 0.689  & 0.25& -0.095 \\
   &&       &      &        &       & Ti(2e) & 0.5 & 0.201  & 0.25& \\
\hline
    & special $B19$ (hcp) 
    &   & 1.732  & 1.633 & 90    & Pd(2e)& 0&0.6666&0.25\\
   &&       &       &       &       &Ti(2e)& 0.5 & 0.1666 & 0.25 \\
    & special $B19$ ($B2$) 
    &   & 1.414  & 1.414 & 90    &Pd(2e)& 0 &0.5&0.25\\
   &&       &       &       &       &Ti(2e)& 0.5 & 0.0 & 0.25 \\
    \hline
    PtTi & $B19$      & 29.84 & 1.758 & 1.668 & 90    & Pt(2e) & 0   & 0.6874 & 0.25& -0.1512\\
    &&       &       &       &       & Ti(2e) & 0.5 & 0.1958 & 0.25 \\
     & $B19'$ 
     & 29.88 & 1.762 & 1.656 & 93.56 & Pt(2e) &0.0154&0.6841 & 0.25& -0.1561\\
    &&       &       &       &       & Ti(2e) &0.4455&0.1923 & 0.25 \\
    & $L1_0$ 
     & 29.52 & 1.000 & 1.386 & 90    &  &    &   &   & -0.1397\\
    & bco
    & 30.57 & 1.835 & 1.561 & 105.80 & Pt(2e) &0.0907&0.6816 & 0.25& -0.1231\\
   &&       &       &       &       & Ti(2e) &0.3502&0.2001 & 0.25 \\
     & $B19$   \footnotemark[1]
     & 30.66  & 1.75  & 1.663  & 90    &&&&\\
     & $B19$   \footnotemark[2]
     & 29.75  & 1.75  & 1.666  & 90    &&&&\\
     & $B19$   \footnotemark[3]
     &  30.9  & 1.72  & 1.62   & 90   & Pt(2e) & 0   & 0.688  & 0.25& -0.155 \\
    &&        &       &        &       & Ti(2e) & 0.5 & 0.197  & 0.25 \\
    \end{tabular}
    \end{ruledtabular}
    \footnotetext[1]{X-ray, from Ref.\ \onlinecite{dwight}.}
    \footnotetext[2]{X-ray, from Ref.\ \onlinecite{donkersloot}.}
    \footnotetext[3]{previous calculation, from Ref.\ \onlinecite{Ye:1997}.}
\end{table*}

In the $B2$ structure, a doubly-degenerate unstable $M_5'$ zone-boundary mode 
implies a doubling of the unit cell to {\bf a}=$(001)_{bcc}$,  {\bf b}=$(110)_{bcc}$, 
{\bf c}=$(1\overline{1}0)_{bcc}$. For the distortion to be frozen in, we choose 
the eigenvector that gives the space group $Pmma$ of the $B19$ 
structure (Figure~\ref{pattern}). The unit cell is orthorhombic, with lattice 
parameters $a$, $b$ and $c$, two Wyckoff positions occupied by Ti and Pd/Pt yielding 
a total of two free internal parameters $v_{Pd/Pt}$ and $v_{Ti}$.
The unstable mode corresponds to a distortion with fixed $v_{Pd/Pt}/v_{Ti}$ 
ratio. In PdTi, the ratios for the phonon and unstable force constant eigenmode as 
computed by PWSCF (VASP) are 
1.55 (1.48) and 1.75 (1.64), respectively. In PtTi, the corresponding ratios are
1.25 (1.19) and 1.81 (1.62), respectively.

In Table~\ref{table:struct}, we provide the calculated equilibrium lattice parameters
of the $B19$ structure, obtained by relaxing all five free structural parameters.
The results are in good agreement with experiment, aside from the volume underestimate 
typical of LDA. The computed relaxed values of the ratio $v_{Pd/Pt}/v_{Ti}$ are 1.29 
for PdTi and 1.16 for PtTi. The latter is quite close to the VASP phonon result, 
while the former is significantly lower.
To understand this result more fully, we separate the coupling to strain
from that to the second (stable) mode of the same symmetry by relaxing
the structure to $B19$ with the lattice held fixed. The resulting ratios 
$v_{Pd/Pt}/v_{Ti}$ are 1.51 for PdTi and 1.41 for PtTi. 
Thus both couplings are significant in both systems, though the effects 
fortutiously nearly cancel in PtTi.
This behavior is compatible with continuum models\cite{james}.

Comparing the parameter values in Table~\ref{table:struct}, we see that the relaxed $B19$ 
structure is close to the ideal hcp structure that would
be obtained from packing monodisperse hard spheres,
with parameters given in the last line of the Table~\ref{table:struct}.
The lower symmetry arises from the ``decoration'' of the close packed plane with 
two different atomic species, which cannot preserve symmetry and maximise unlike 
near neighbors\cite{Kelsey}. This confirms that the
phase transition is best thought of as a binary equivalent of bcc-hcp,
not simply as a distortion of the $B2$ structure.  There is an
interesting contrast with NiTi here: the experimentally reported
ground state NiTi $B19'$ phase has $b/a=1.603$ and
$c/a=1.417$\cite{KTMO}, far from hcp.  

\begin{table}[htbp]
    \caption{
     \label{table:phonon-gamma}
     Optical phonon frequencies at $\Gamma$ for PdTi and PtTi 
     in the $B19$ and bco structures by DFPT.  
     The lowest frequency mode couples to the strain in the $B19$-$B19'$ transition.}
    \begin{ruledtabular}
    \begin{tabular}{ccc}
                     Alloy & structure & $\omega$ (cm$^{-1}$) \\
    \hline
    PdTi & $B19$ & 73, 90, 101, 139, 159, 184, 199, 231, 251 \\
         & bco & 94, 138, 146, 148, 164, 198, 200, 230, 256 \\
    PtTi & $B19$ & 68, 87, 98, 123, 165, 185, 196, 238, 260 \\
         & bco & 119, 171, 172, 173, 185, 210, 225, 247, 284 \\
    \end{tabular}
    \end{ruledtabular}
\end{table}

\begin{table}
    \caption{\label{table:ratio}
     Calculated optical phonon frequencies and ratio of atomic
     displacements of $\Gamma_{4}'$ mode of PdTi and PtTi at relaxed $B19$
     structure from force constant matrix eigenvectors (i.e. ignoring masses).
     The values in italic are the modes leading to $B19'$ structure and are 
     compared with values in parentheses
     which are taken from relaxed $B19'$ structure.  Frequencies here differ 
     slightly from  Table~\protect\ref{table:phonon-gamma} because they are calculated
     from finite displacements\protect\cite{prac} using VASP rather than DFPT.}
    \begin{ruledtabular}
    \begin{tabular}{ccc}
     Alloy & Frequencies & $u_{Pd/Pt}/u_{Ti}$ \\
    \hline
     PdTi & 74, 103, 159 & {\it -0.280} (-0.217), 3.5684, -1.002 \\
     PtTi & 69, 106, 161 & {\it -0.321} (-0.283), 3.1159, -1.05\\
    \end{tabular}
    \end{ruledtabular}
\end{table}

\begin{figure*}
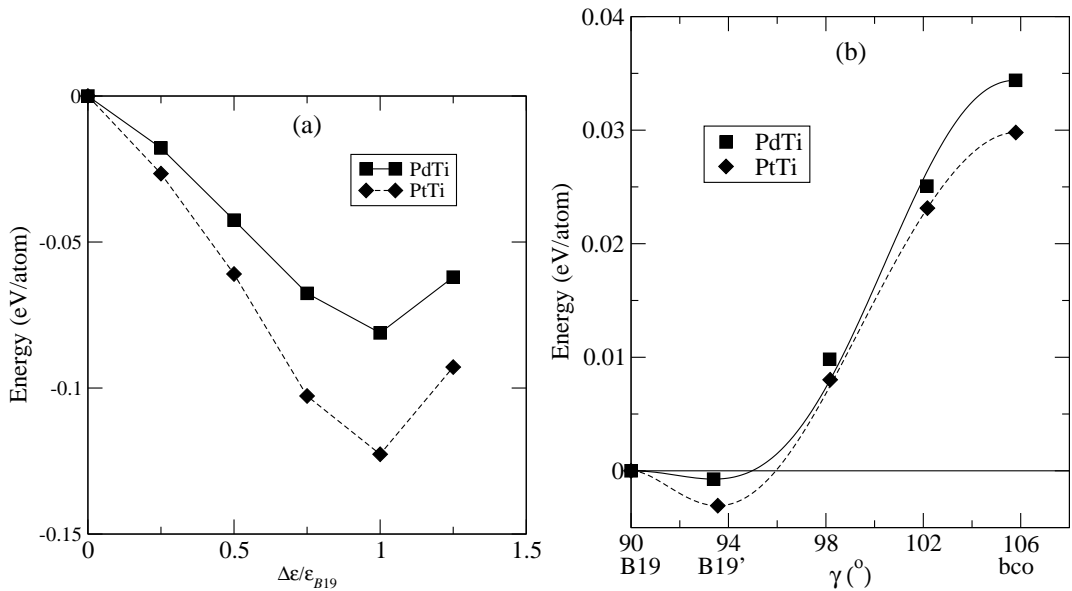

\includegraphics[scale=0.45]{fig-3a.eps}
\includegraphics[scale=0.45]{fig-3b.eps}
  \caption{\label{tote}
(a) Total energies as a function of uniaxial strain with the 
internal coordinates being fully relaxed. The lattice parameters 
for $\Delta\varepsilon/\varepsilon_{B19}$=0, 1 are 
taken from $B2$ and $B19$ structures, while others were fixed by 
interpolation.  The ``$B2$'' structure here corresponds to freezing in the 
$M$-point phonon, and lies 10.5~meV/atom (PdTi) or 30.2~meV/atom (PtTi)
below the fully symmetric ($v=0$) structures.
(b)Total energies as a function of monoclinic angle. For
$B19$, $B19'$ and bco all degrees of freedom are relaxed.  For intermediate
angles the lattice parameters $a$, $b$, $c$ were fixed by interpolation, 
while the internal coordinates $u$, $v$ were fully relaxed. $u$ and $v$ vary 
almost linearly with $\gamma$.  Symbols indicate calculated values, curve is 
a polynomial fit with $dE/d\gamma=0$ enforced where required by symmetry.}
\end{figure*}

In the soft-mode approach, there is no guarantee that the
energy-minimizing freezing-in of one unstable
mode will stablize the other unstable modes of the high-symmetry
structure.  In the present case, the undistorted doubled unit cell 
contains two sets of 
(110)$_{bcc}$ planes each of which is unstable to strain-coupled
shuffling at all {\it q-}points. 
The simplest such mode in the $B19$ structure is $\Gamma_4$, which 
lowers the symmetry to monoclinic $P2_1/m$.
The resulting $B19'$ structure has three additional free parameters: 
the monoclinic angle $\gamma$ and two internal parameters $u_{Ti}$ and $u_{Pd}$.
These values are given for the relaxed $B19'$ structure in Table~\ref{table:struct},
and compared with the normalized eigenvector displacements in Table~\ref{table:ratio}.
However, it is important to note that the computed $\Gamma$ phonon modes in the $B19$ 
structures are in fact all stable (Table~\ref{table:phonon-gamma}). 
The lowering of energy by distortion to $B19'$ cannot be obtained
by a pure $\Gamma_4$ phonon distortion, but only if the strain is allowed to
relax simulataneously (Table~\ref{table:struct}).
This may be the reason that in a previous calculation \cite{Ye:1997}, $B19$ was
reported to be the minimum energy structure.

The relative energies of the various relaxed structures
are given in Table \ref{table:struct}.
The $B19$ total energy is lower than $B2$.  
A simple estimate of the transition temperature is given by
$\Delta E = kT_c$ which suggest $T_c$ of 
1050K (PdTi) and 1755K (PtTi).
These rough values are significantly larger than
the experimental data for the hysteretical transition 
region\cite{donkersloot,data1,data2} 
(approximately 800K and 1400K respectively),
but show the correct material trend.
For neither system has a $B19'$ phase yet been 
observed experimentally. 
The small computed energy differences between $B19'$ and
$B19'$, translated into temperature, are
28K and 39K for PdTi and PtTi, respectively.
This suggests that the transition to the $B19'$ phase
should occur at temperatures well below those at which the 
experiments were performed, so that our results are 
fully consistent with the available experimental work.  

\begin{figure*}
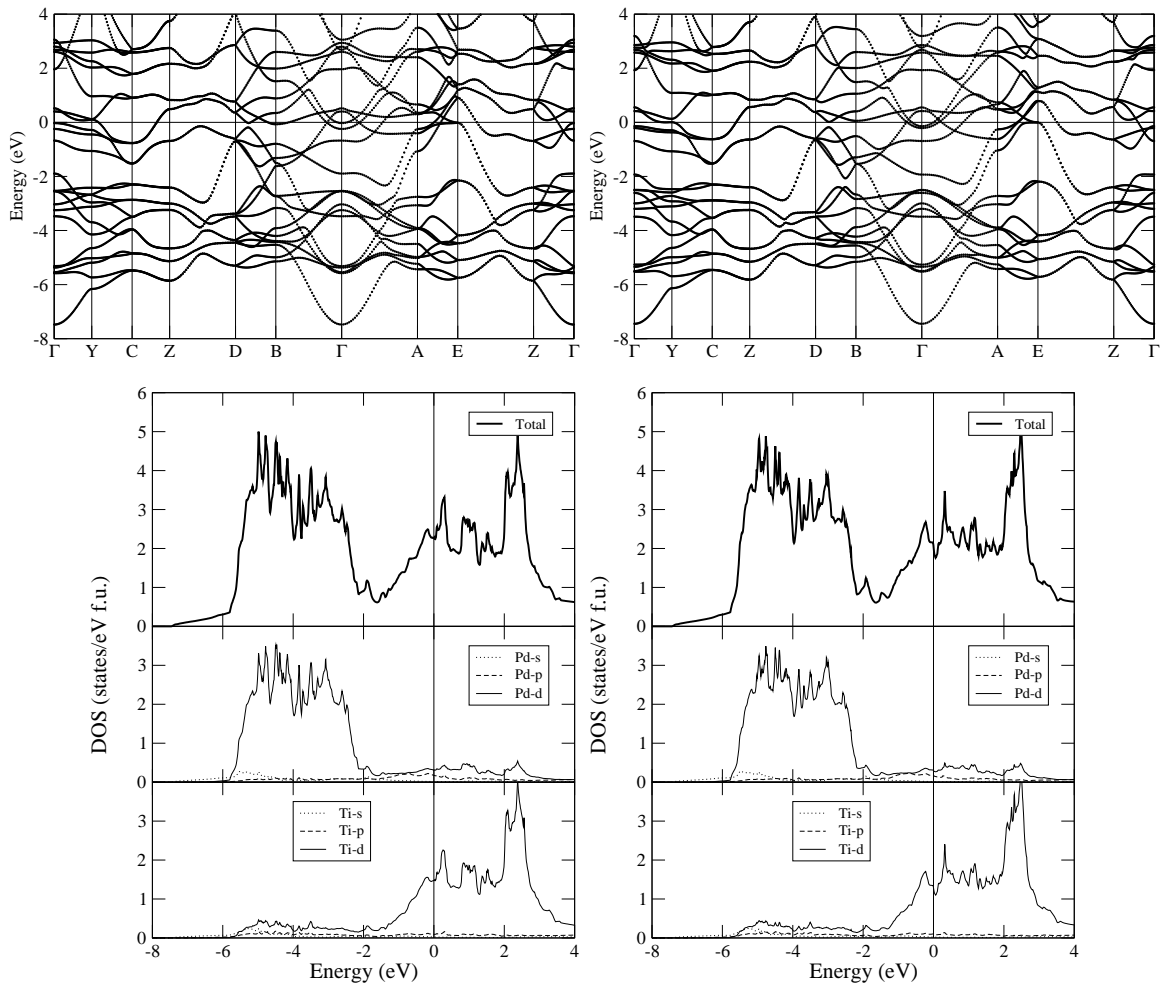

\includegraphics[scale=0.4]{fig-4a.eps}
\includegraphics[scale=0.4]{fig-4b.eps}\\[0.3cm]
\includegraphics[scale=0.4]{fig-4c.eps}
\includegraphics[scale=0.4]{fig-4d.eps}
  \caption{\label{bands-pdti}
Band structures, densities of states and partial 
densities of states of PdTi at relaxed $B19$ (left) and 
equilibrium $B19'$ (right) structures.
Labelling of the $B19$ band structure is as a special case of $B19'$.}
\end{figure*}

\begin{figure*}
\includegraphics[scale=0.4]{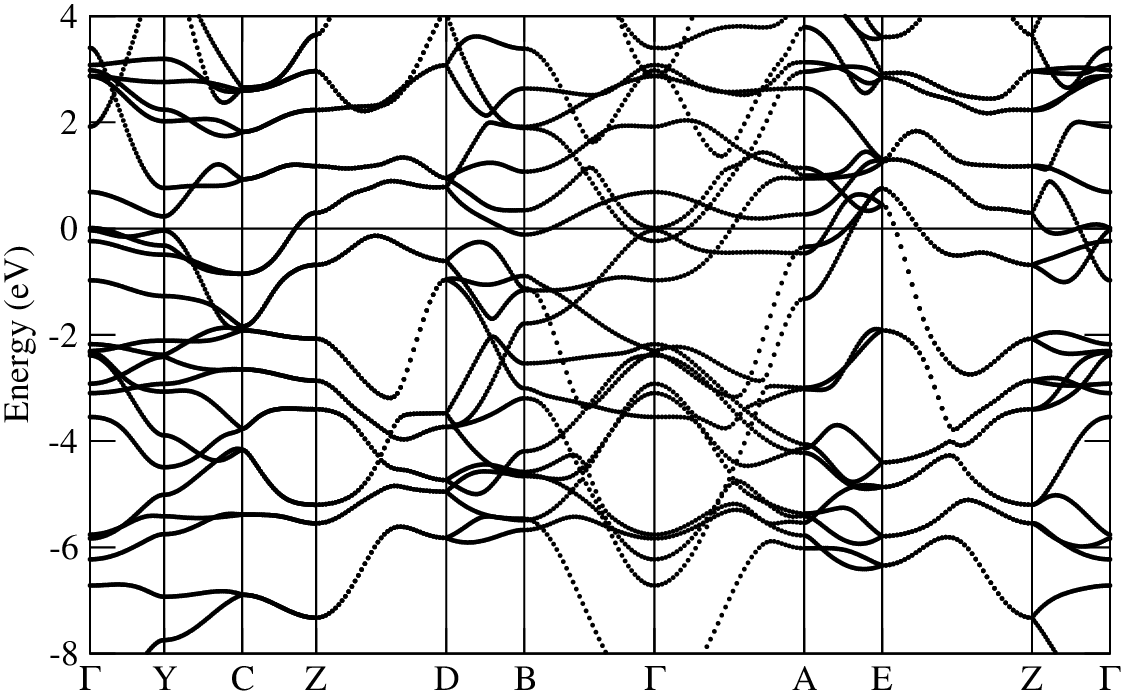} 
\includegraphics[scale=0.4]{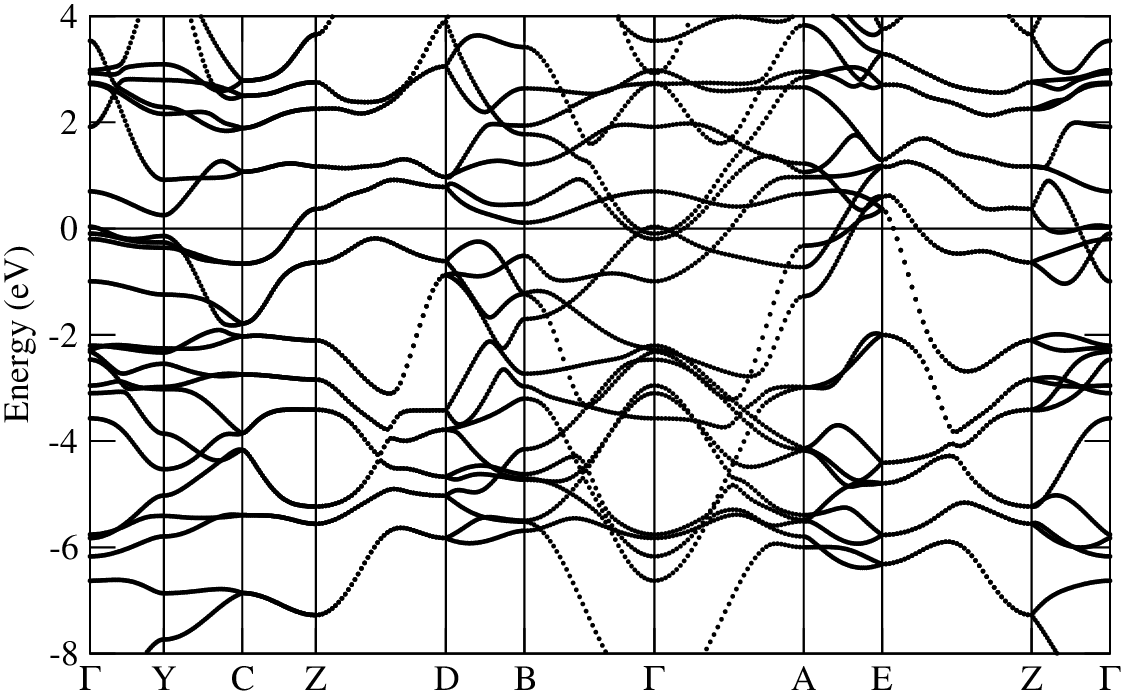}\\[0.3cm]
\includegraphics[scale=0.4]{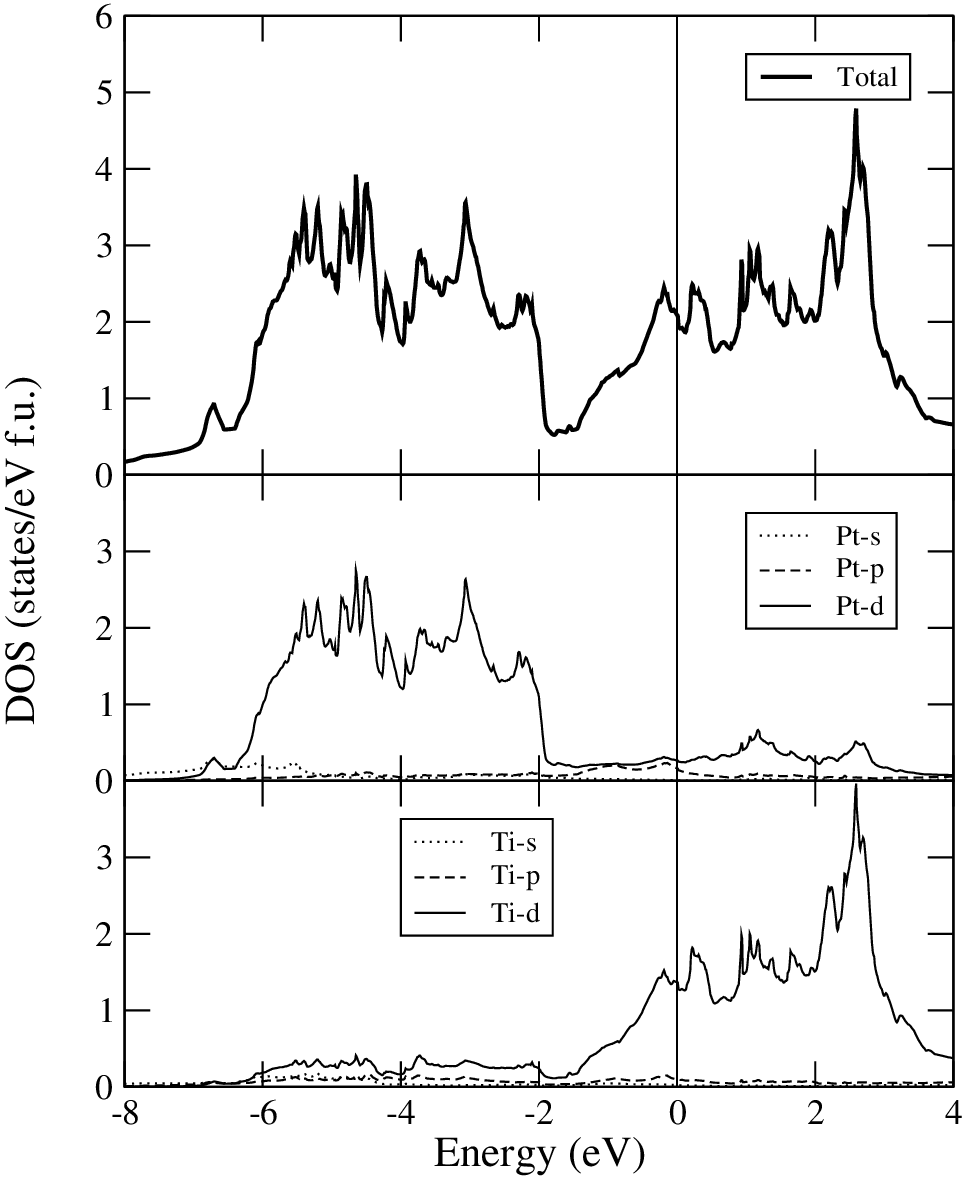} 
\includegraphics[scale=0.4]{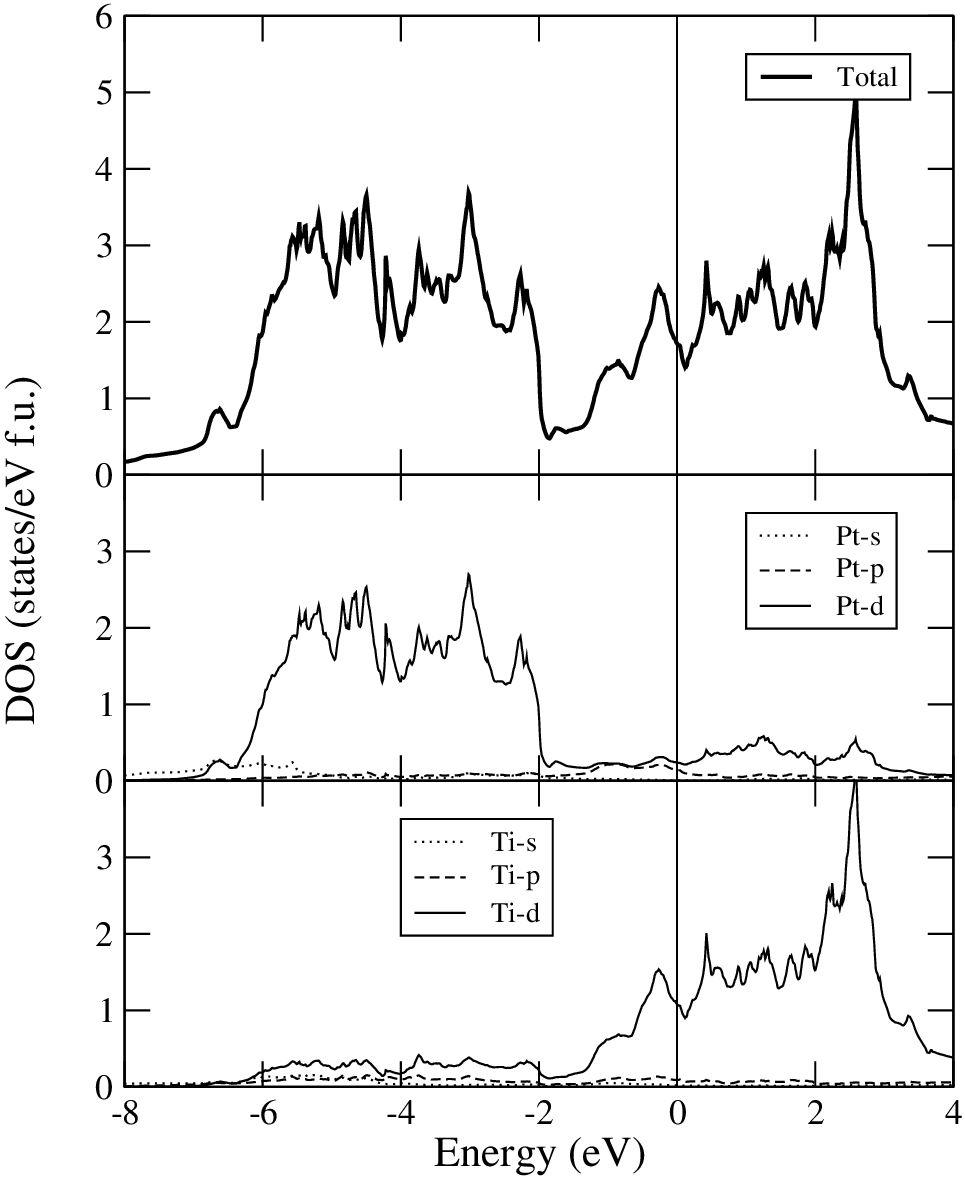}
  \caption{\label{bands-ptti}
Band structures, densities of states and partial 
densities of states of PtTi at relaxed $B19$ (left) and 
equilibrium $B19'$ (right) structures.
Labelling of the $B19$ band structure is as a special case of $B19'$.}
\end{figure*}

The binary-hcp phase interpretation also
suggests that we should examine the binary-fcc equivalent, which is the
$L1_0$ phase.  $L1_0$ is accessible from $B2$ by a simple (001) shear
and has lower energy\cite{BEN}, however we find that $L1_0$ has
slightly higher energy than $B19$.
We consider one further structure: at the special $B19'$ 
values $\cos\gamma = a/2b$ and $4u-1=2v$, orthorhombic symmetry 
is restored. The side of the conventional cell of this body centered orthorhombic 
(bco) (space group $Pmna$)
structure is doubled in the {\bf b} direction, 
though the primitive cell still contains four atoms.  Although 
at relatively high energy (~\ref{table:struct}), 
this provides us a reference point 
for structures with large $\gamma$\cite{niti}.  Note that a 
further shear to $\cos\gamma = a/b$ would give the $B19$
structure once more.

The potential energies of continuous paths between the structures identified above 
are important for understanding the transformation mechanism.
We compute the energies of three paths: $B2$-$B19$, $B19$-$B19'$ and 
$B19'$-bco. 
In keeping with the timescale separation between bulk strain and 
atomic motion, we define intermediate configurations by relaxing 
the atoms to their minimum energy configuration consistent with the
applied symmetry and strain on the cell.  The remaining four 
strain degrees of freedom are reduced to a single parameter by taking
interpolations between the strains of the endpoint structures. For the
$B2$ structure, we minimize the energy assuming the $B19$ space
group, which gives a smooth evolution of the structure along the path.
>From Figure~\ref{tote}, it is clear that there is no total energy barrier along the
$B2$-$B19$-$B19'$ path, and that $B19$ represents the total energy
barrier between $B19'$ martensitic variants.
The $B19$ phase can be viewed as the binary equivalent of hcp, and the
transformation path as the binary equivalent of the Nishiyama-Wassermann (NW) path. 
Using the analogy with the NW mechanism for the
bcc-hcp transition, we can attribute the transition to a shuffling of 
(110)$_{bcc}$ planes\cite{Pinsook}.


Structural instabilities in metals are typically related to details 
of the Fermi surface, and we have calculated the band structures of $B2$, 
$B19$ and $B19'$ to investigate this.  
In PdTi/PtTi the band structure is dominated 
by the {\it d}-bands, with the Pd/Pt bands lying below the Fermi level
and almost fully occupied,
and the Ti $d$-bands lying above the Fermi level (Figure~\ref{bands-pdti} 
and ~\ref{bands-ptti}), the band centers being offset by some 6eV.
The free-electron like $sp$ bands are very broad, and play little 
role in the bonding except to donate some electrons to the Ti-d band.
The large strain involved in the $B2-B19$ transition means that the 
Fermi surfaces are quite different ($B19$ having the lowest DOS at 
$E_F$\cite{Ye:1997}) and this transition cannot be related directly 
to the band structure.
By contrast, the $B19$-$B19'$ transition is accompanied by the opening of a 
pseudogap at the Fermi level, a typical signature of increased 
stability.  The band structures are very similar, the small
difference which stabilises $B19'$ being traceable to the shifting 
above the Fermi level of a pocket of electrons around $B$. The distortion 
to $B19'$ is just enough to complete the topological phase transition 
which eliminates this pocket of electrons in both materials.

\section{Conclusions}

In conclusion, 
we have performed {\it ab initio} calculations of the structural energetics of 
PdTi and PtTi.
In each case we predict that the low temperature ground state structure 
will be $B19'$, with the (observed) $B2$ and $B19$ phases being dynamically 
stabilized.
There are no total energy barriers between the structures, meaning that the
phase space microstates that belong to the $B19'$ structure {\it also} 
belong to the $B19$ and $B2$ structures.

In contrast to NiTi\cite{Huang:2002}, the entire $\Gamma-M$ phonon 
branch is unstable. We showed that the $B19$ structure can be obtained by 
a ``freezing in'' of phonons of the $B2$ structure coupled to the shear
associated with the $c'=(c_{11}-c_{12})/2$ elastic constant, but that 
no single dynamical-matrix or force-constant-matrix eigenvector leads
to the low-symmetry phase.

The $B19'$ then corresponds to a further strain 
coupled to a $\Gamma_4$ phonon of the $B19$ 
phase.  Tracing the atomic motions of these instabilities shows that they 
are both related to shears of alternate $(110)_{B2}$ phases, and hence 
that the transition mechanism is the binary equivalent of the 
Nishiyama-Wassermann bcc-hcp mechanism.


\acknowledgments

We thank R. D. James, I. I. Naumov, and K. Bhattacharya
for valuable discussions. This work was supported by AFOSR/MURI F49620-98-1-0433.
The calculations were performed on the SGI Origin 3000 and IBM SP3 at 
ARL MSRC.


\begin{thebibliography}{1}
\bibitem{niti} According to various density functional calculations, 
this body centered orthorhombic structure appears to be more stable 
than $B19'$ in stoichiometric NiTi.
\end{thebibliography}
\end{document}